\documentclass[sigconf]{acmart}
\usepackage{url,subfigure,ulem}
\usepackage{array}

\usepackage{algorithm,algorithmic,paralist}
\usepackage{listings}
\usepackage{color}
\usepackage{graphicx}
\usepackage{hyperref}
\usepackage{multirow}
\usepackage{threeparttable}
\AtBeginDocument{%
  \providecommand\BibTeX{{%
    \normalfont B\kern-0.5em{\scshape i\kern-0.25em b}\kern-0.8em\TeX}}}

\begin{document}

\title{Gaussian Graph with Prototypical Contrastive Learning in E-Commerce Bundle Recommendation}

\author{Zhao-Yang Liu}
\authornote{Both authors contributed equally to this research.}
\orcid{1234-5678-9012}
\author{Liucheng Sun}
\authornotemark[1]
\email{{shuqian.lzy,liucheng.slc}@alibaba-inc.com}
\affiliation{%
  \institution{Alibaba Group}
  \city{Hangzhou}
  \state{Zhejiang}
  \country{China}
}

\author{Chenwei Weng}
\email{wengchenwei.pt@alibaba-inc.com}
\affiliation{%
  \institution{Alibaba Group}
  \city{Hangzhou}
  \state{Zhejiang}
  \country{China}
}

\author{Qijin Chen}
\email{qijin.cqj@alibaba-inc.com}
\affiliation{%
  \institution{Alibaba Group}
  \city{Hangzhou}
  \state{Zhejiang}
  \country{China}
}

\author{Chengfu Huo}
\email{chengfu.huocf@alibaba-inc.com}
\affiliation{%
  \institution{Alibaba Group}
  \city{Hangzhou}
  \state{Zhejiang}
  \country{China}
}




\begin{abstract}
Bundle recommendation aims to provide a bundle of items to satisfy the user preference on e-commerce platform. Existing successful solutions are based on the contrastive graph learning paradigm where graph neural networks (GNNs) are employed to learn representations from user-level and bundle-level graph views with a contrastive learning module to enhance the cooperative association between different views. Nevertheless, they ignore the uncertainty issue which has a significant impact in real bundle recommendation scenarios due to the lack of discriminative information caused by highly sparsity or diversity. We further suggest that their instance-wise contrastive learning fails to distinguish the semantically similar negatives (i.e., sampling bias issue), resulting in performance degradation.

In this paper, we propose a novel \textbf{G}aussian Graph with \textbf{P}rototypi-
cal \textbf{C}ontrastive \textbf{L}earning (GPCL) framework to overcome these challenges. In particular, GPCL embeds each user/bundle/item as a Gaussian distribution rather than a fixed vector. We further design a prototypical contrastive learning module to capture the contextual information and mitigate the sampling bias issue. Extensive experiments demonstrate that benefiting from the proposed components, we achieve new state-of-the-art performance compared to previous methods on several public datasets. Moreover, GPCL has been deployed on real-world e-commerce platform and achieved substantial improvements.
\end{abstract}






\maketitle

\section{Introduction}
\begin{figure}
  \includegraphics[width=0.8 \columnwidth,height=0.42\linewidth]{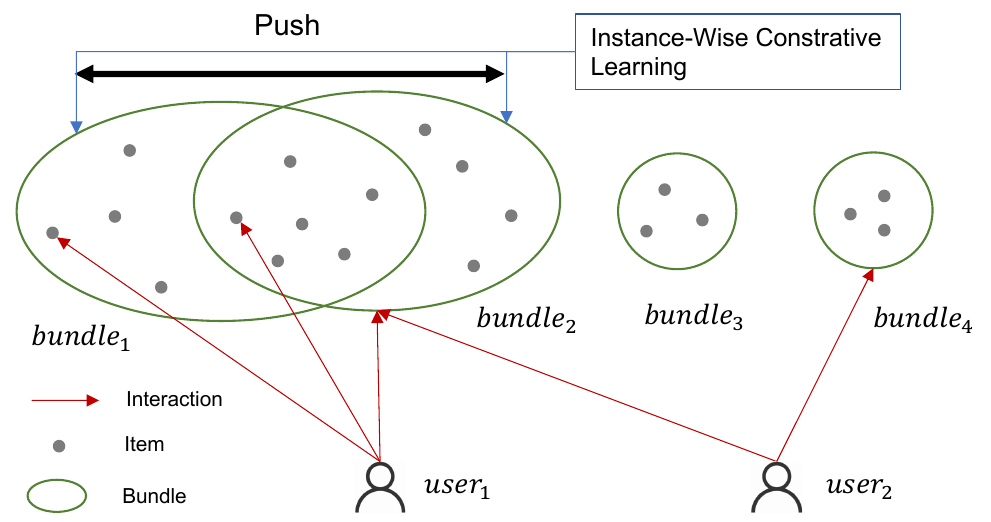}
  \caption{Illustration of user-bundle-item relations.}
  \label{case}
\end{figure}
Recommenders are ubiquitous — they're in many of the apps and tools we use every day. Recommender systems serve users with 
personalized items and handle the increasing online information overload problem. In recent years, recommendation has evolved into a new form, called bundle recommendation \cite{steam, br14}, where a collection of items are recommended as a whole (e.g. music playlist and travel package). In e-commerce, item bundling \cite{tao22,item-tag} is an explainable and widely-used representation of user interests, and provide an effective strategy to support promotional campaigns (e.g., purchasing a bundle has a discounted rate). 
As shown in Figure \ref{case}, users can directly interact with items and bundles.
, and a bundle is usually composed of items with identical or similar attributes.

Developing effective bundle recommendation system algorithms has received significant attention by scholars and practitioners\cite{bnet, bgcn,dam, cbr}. The central issue of the problem is to model a user's preference for a bundle. As pointed out in DAM \cite{dam}, simply adopting the collaborative filtering methods is technically feasible but does not work well due to the non atomic property of bundles and highly sparse interactions. The contribution of content-based models is also limited because there is usually no informative bundle-level content features as item-level (e.g., category and brand), making bundle recommendation a non-trivial task. Therefore existing research \cite{bgcn, cbr} recognizes that representation learning plays a pivotal role in the bundle recommendation tasks, and focuses on how to fully take advantage of the user-bundle-item co-occurrence information. Specifically, the user/bundle representations can be inferred from two views \cite{bgcn}: 
\begin{itemize}
\item \textbf{Bundle view} where the user-bundle interactions are emphasized. For example, as shown in Figure \ref{case}, $user_1$ is likely to prefer $bundle_4$ because the behaviorally similar user $user_2$ prefers $bundle_4$.
\item \textbf{Item view} where we utilize the user-item interactions and the bundle-item affiliation information, e.g., $user_1$ may prefer $bundle_1$ because she has been showing interest in $bundle_1$'s affiliated items. 
\end{itemize}
In a major advance in 2022, CrossCBR \cite{cbr} model the two views separately to perform representation learning, and then enhance the view-aware representations by fusing them in the following joint optimization stage with the cross-view contrastive and BPR \cite{bpr} loss, achieving state-of-the-art (SOTA) performance.

Despite their success, the previous work suffers from the following limitations. Firstly, existing approaches \cite{icl-4,icl-5,icl-6} mainly focus on capturing the instance-level similarity but fail to discover the underlying structure contexts over the whole data distribution. Considering the running example in Figure \ref{case}, $bundle_1$ and $bundle_2$ share lots of items in common, thus the distance between their vectors might be close intuitively. However, the representation in traditional instance-wise contrast learning \cite{cbr,icl-1,icl-2,icl-3} is not encouraged to encode this contextual semantic structure of data. Similar bundles may be pushed away with random negative sampling. This phenomenon, which we call “sampling bias”, can empirically lead to significant performance degradation \cite{pgcl}. Secondly, the existing studies \cite{dam,bnet} represent each node by a single point in a low-dimensional continuous vector space but fail to incorporate uncertainty which is caused by the lack of discriminative information. We claim that the uncertainty is inherent when describing a node by a single point in the real-world e-commerce bundle recommendation. In practice, a bundle may be composed of a wide range of genres of items (e.g. $bundle_1/bundle_2$ in Figure \ref{case}), and very limited information could be learned from its interactions since it is not representative enough. In addition, a large number of bundles/users with very few interactions are difficult to represent precisely due to the highly sparsity.

Towards this end, we propose a \textit{\textbf{G}aussian Graph with \textbf{P}rototypical \textbf{C}ontrastive \textbf{L}earning} (GPCL) framework to incorporate the uncertainty and discover the contextual semantic structure of data. In particular, unlike the prior approaches that represent an instance as a single point vector in a low-dimensional continuous space, we treat every instance as a Gaussian distribution consisting of a mean and a variance vector to describe uncertainty. Then a group of samples from the distribution are generated through Monte-Carlo sampling as the inputs of subsequent components. In addition, we design a prototypical contrastive learning module, where a prototype is defined as a representative embedding for a group of semantically similar nodes. By allocating the similar nodes to a learned prototype, we then pull together positive pairs of $\langle$node, prototype$\rangle$ and push away negative ones instead of pairs of $\langle$node, node$\rangle$. In this way, the contextual semantic similarity is captured. At the time of writing, the method was fully deployed on a large e-commerce platform for the bundle recommendation task. In short, the main contributions are summarized as follows.
\begin{enumerate}
    \item We present a model architecture agnostic Gaussian embedding module to capture uncertainty and enhance model representation capability. Learning uncertainty of embedding produces meaningful representations for modeling the user preference, and it also enables the downstream exploration and exploitation strategy in the recommendation systems.
    \item A novel prototypical contrastive learning is developed to learn node representations better and further in a self-supervised manner, which can alleviate the "sampling bias" issue in the instance-wise contrastive learning for bundle recommendation.
    \item Extensive offline experiments demonstrate the state-of-the-art performance of our method and its effectiveness of each component. We also exhibit the superiority of our method in a real-world e-commerce production environment.
\end{enumerate}


\section{PRELIMINARIES}
Before presenting our methodology, we first formally define the problem and introduce the base model.

\subsection{Problem Definition}
In order to integrate item level information to improve bundle recommendation accuracy, there are two types of important side information that need to be modeled, the user’s preference to item and bundle’s composition information.
Let $\boldmath{U}=\{u_i|i=1,2,...,M\}$, $\boldmath{I}=\{i_j|j=1,2,....N\}$ and $\boldmath{B}=\{b_k|k=1,2,...,O\}$ denote the sets of users, items and bundles respectively, where $M,N,O$ is the sizes of corresponding sets. We define user-bundle and user-item interaction matrix as $X_{M\times O}=\{x_{ub}|u\in U\}$ and $Y_{M\times N}=\{y_{ui}|u\in U, i\in I\}$. 
The bundle-item affiliation matrix is represented by $Z_{O\times N}=\{z_{bi}|b\in B, i\in I\}$. We define the binary variables $x_{ub}, y_{ui}, z_{bi}$ that assume value 1 if and only if user $u$ interacts with bundle $b$ or user $u$ interacts with item $i$, or item $i$ belongs to bundle $b$. As illustrated in Section 1, users can directly click/purchase/collect items or interact with bundles, thus X and Y are independent. Based on the historical interactions and affiliation relations, our goal is to estimate the probability of the user $u$ interacting with bundle $b$.

\subsection{Base Model}
Due to the success of cross-view contrastive learning for bundle recommendation in CrossCBR \cite{cbr}, we take it as the base model and also a part of the GPCL as shown in Figure \ref{over}. The key idea is that the bundle-view and item-view provide complementary but different information about the user-bundle preference, thus it is worthy to perform representation learning upon two views separately and further design a contrastive loss (CL) to model the cooperative association and achieve mutual enhancement.

For the bundle-view, an undirected bipartite user-bundle (U-B) graph is constructed based on the historical interaction matrix $X$. Then the prevailing GNN-based recommendation framework LightGCN \cite{lgcn} is adopted as the encoder to learn the representations of both users and bundles. For the simplicity of notations, assume that the number of LightGCN layers takes 1, 
$e_u^0$ and $e_b^0$ are the initialized node vectors of the user $u$ and bundle $b$, then the final representations are denoted as:
\begin{equation}
\begin{split}
&e_u^B=\sum_{b\in N_u}\frac{1}{\sqrt{|N_u|}\sqrt{|N_b|}}e_b^0\\
&e_b^B=\sum_{u\in N_b}\frac{1}{\sqrt{|N_b|}\sqrt{|N_u|}}e_u^0
\end{split}
\end{equation}
 where $N_u$ and $N_b$ are the neighbor nodes of the user $u$ and bundle $b$ respectively.

For the item-view, user-item (U-I) graph and bundle-item (B-I) graph are constructed according to the user-item interactions $Y$ and bundle-item affiliations $Z$, respectively. Similar to the U-B graph learning, $e_i^0$ and $e_u^0$ are the initialized node vectors of the item $i$ and user $u$, based on U-I graph we have:
\begin{equation}
\begin{split}
&e_u^I=\sum_{i\in N_u}\frac{1}{\sqrt{|N_u|}\sqrt{|N_i|}}e_i^0\\
&e_i^I=\sum_{u\in N_i}\frac{1}{\sqrt{|N_i|}\sqrt{|N_u|}}e_u^0
\end{split}
\end{equation}
where $N_i$ is the neighbor nodes of item $i$. According to the B-I graph, the neighborhood of the bundle is aggregated by average pooling to get the representation of bundle $b$ as:
\begin{equation}
e_b^I=\frac{1}{|N_b|\sum_{i\in N_b}e_i^I}
\end{equation}

The popular InfoNCE \cite{info-nce} loss is then built upon the cross-view representations of users and bundles, respectively. More precisely, the contrastive loss is able to simultaneously encourage the alignment of the same user/bundle from different views and enforce the separation of different users/bundles. The contrastive loss $L^U_{CL}$ and $L^B_{CL}$ are as follows:
\begin{equation}
\begin{split}
&L^U_{CL}=\frac{1}{|U|}
\sum_{u\in U}-\text{log}\frac{\text{exp}(e_u^B\cdot e_u^I)/\tau)}{\sum_{v\in U}\text{exp}((e_u^B\cdot e_v^I)/\tau)}\\
&L^B_{CL}=\frac{1}{|B|}
\sum_{b\in B}-\text{log}\frac{\text{exp}(e_b^B\cdot e_b^I)/\tau)}{\sum_{p\in B}\text{exp}((e_b^B\cdot e_p^I)/\tau)}
\end{split}
\end{equation}
where $\tau$ is the hyper-parameter, and $v, p$ are negative samples. By summing the two term, we obtain the final cross-view contrastive loss:
\begin{equation}
L_{CL}=L_{CL}^U+L_{CL}^B
\end{equation}

After we obtain the representations of users and bundles from two views, the prediction is made by:
\begin{equation}
y_{u,b}=e_u^B\cdot e_b^B+e_u^I\cdot e_b^I
\end{equation}

The Bayesian Personalized Ranking (BPR) [14] loss function is adopted as the main loss:
\begin{equation}\label{pred}
L^{BPR}=\sum_{u,b,b'}-\text{ln}\sigma(y_{u,b}-y_{ub'})
\end{equation}

Finally, the overall training loss is:
\begin{equation}
L_{base} = L^{BPR} + \gamma_{CL} * L_{CL}
\end{equation}
where $\gamma_{CL}$ is the weight of the cross-view contrastive loss.
\section{METHODOLOGY}
\begin{figure*}[htbp]
	\begin{center}
		\begin{minipage}{0.9\linewidth}
			\includegraphics[width=\textwidth]{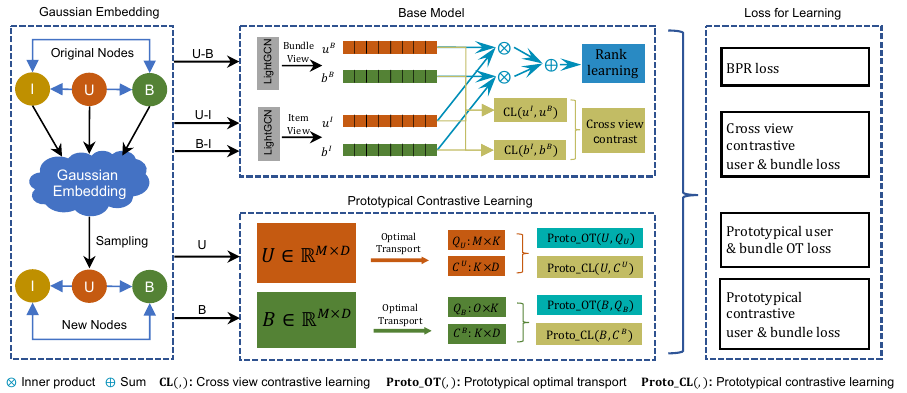}\\
		\end{minipage}
		\caption{Model architecture of the proposed GPCL.}\label{over}
	\end{center}
\end{figure*}

In this section, we first propose two core components: Gaussian embedding module and prototypical contrastive learning module. Then we present the model learning loss and some discussions. The overall architecture of our proposed framework GPCL is shown in Figure \ref{over}.

\subsection{Gaussian Embedding}
We represent each bundle, user or item with an Gaussian distribution governed by a mean vector and a covariance matrix. To limit the complexity of the model and reduce the computational overhead \cite{ijcai19,sigir17}, we assume that the random variable of different dimensions are uncorrelated, thus the covariance matrix $\Sigma$ is diagonal and can be further represented by a variance vector $\sigma$. To be specific, each bundle, user or item has two embedding representations, which are for the mean and variance. 

\begin{figure}[htbp]
  \includegraphics[width=0.85 \columnwidth,height=0.32\linewidth]{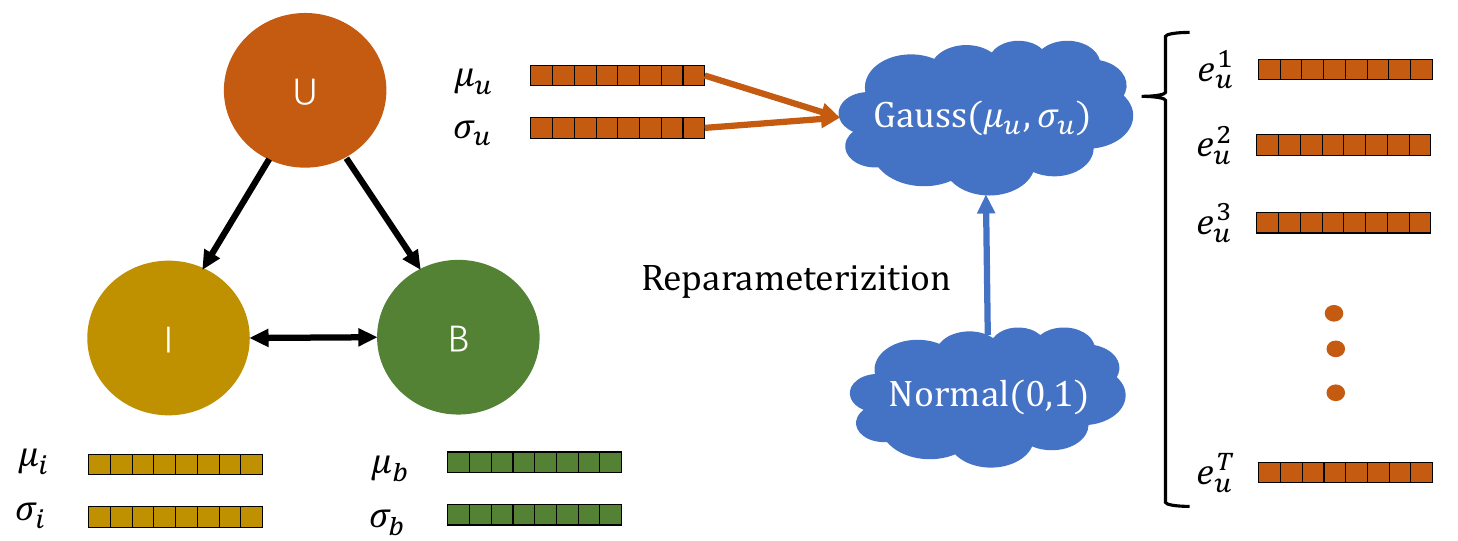}
  \caption{Illustration of Gaussian embedding module.}
  \label{gauss}
\end{figure}

For user $u$, we denote the mean embedding as $\mu_u\in \mathbb{R}^D$ and variance embedding as $\sigma_u \in \mathbb{R}^D$, where $D$ is the embedding dimension. Analogously, we use $\{(\mu_b,\sigma_b)|\mu_b\in \mathbb{R}^D, \sigma_b\in \mathbb{R}^D\}$ and $\{(\mu_i,\sigma_i)|\mu_i\in \mathbb{R}^D, \sigma_i\in \mathbb{R}^D\}$ for user $u$ and item $i$ respectively. In this way, each node is represented by $2\times D$ parameters.

To maintain numerical stability and non-negativity, the variances are transformed by "Exponential Linear Unit" (ELU) \cite{ijcai19}: 
\begin{equation}
    \sigma' =ELU(\sigma)+1
\end{equation}
Then we can get transformed variances: $\sigma'_u$, $\sigma'_b$ and $\sigma'_i$. Note that there are many other activation functions \cite{act}, they are also applicable with appropriate adjustment
 and could achieve similar performances by our test. 

In order to perform back-propagation, we use the reparameterization trick \cite{vae} to obtain the embedding sample as follows:
\begin{equation}
e=\mu+\sqrt{\sigma'} \times \epsilon
\end{equation}
where $\epsilon\sim N(0,1)$. As shown in Figure \ref{gauss}, we can randomly sample several times of users/bundles/items as the input of the subsequent modules. For ease of notation in the following, assume that we only sample once and their Gaussian embeddings are denoted as $e_u$, $e_b$ and $e_i$, respectively.

\subsection{Prototypical Contrastive Learning}
\begin{figure}[htbp]
  \includegraphics[width=0.85 \columnwidth,height=0.44\linewidth]{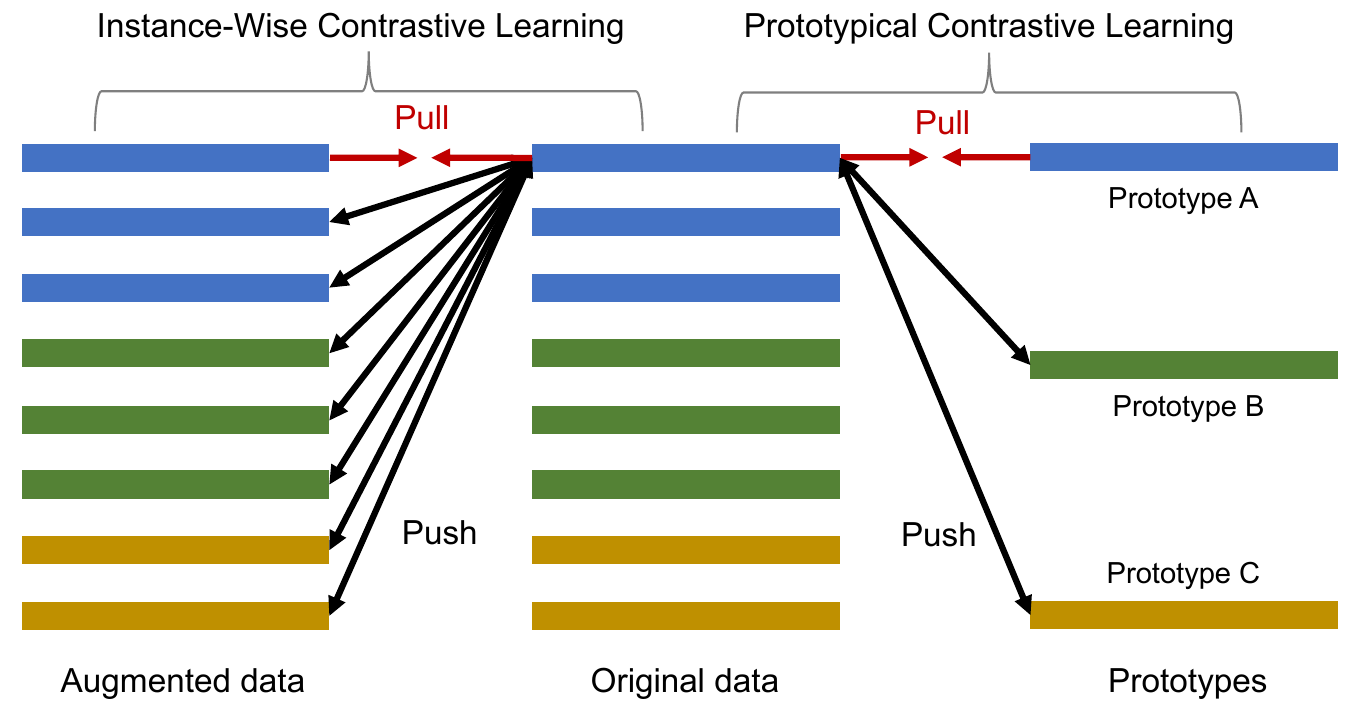}
  \caption{Comparison of Instance-wise and prototypical contrastive learning.}
  \label{cl}
\end{figure}

The cross-view contrastive learning guarantees mutual enhancement of the two views. However, as shown in Figure \ref{cl},
it suffers from the critical "sampling bias" issue that similar nodes might be pushed far apart and leading to performance degradation. To reduce selection bias and obtain a better representation, we design a prototypical contrastive objective to capture the correlations between a user/bundle and its prototype. The prototype can be seen as the context of each user/bundle which represents a group of semantic neighbors even that they are not structurally connected in the graph. Regarding the prototype learning as a type of feature clustering for another view, we can perform prototypical contrastive learning.

\begin{figure}
  \includegraphics[width=0.93 \columnwidth,height=0.5\linewidth]{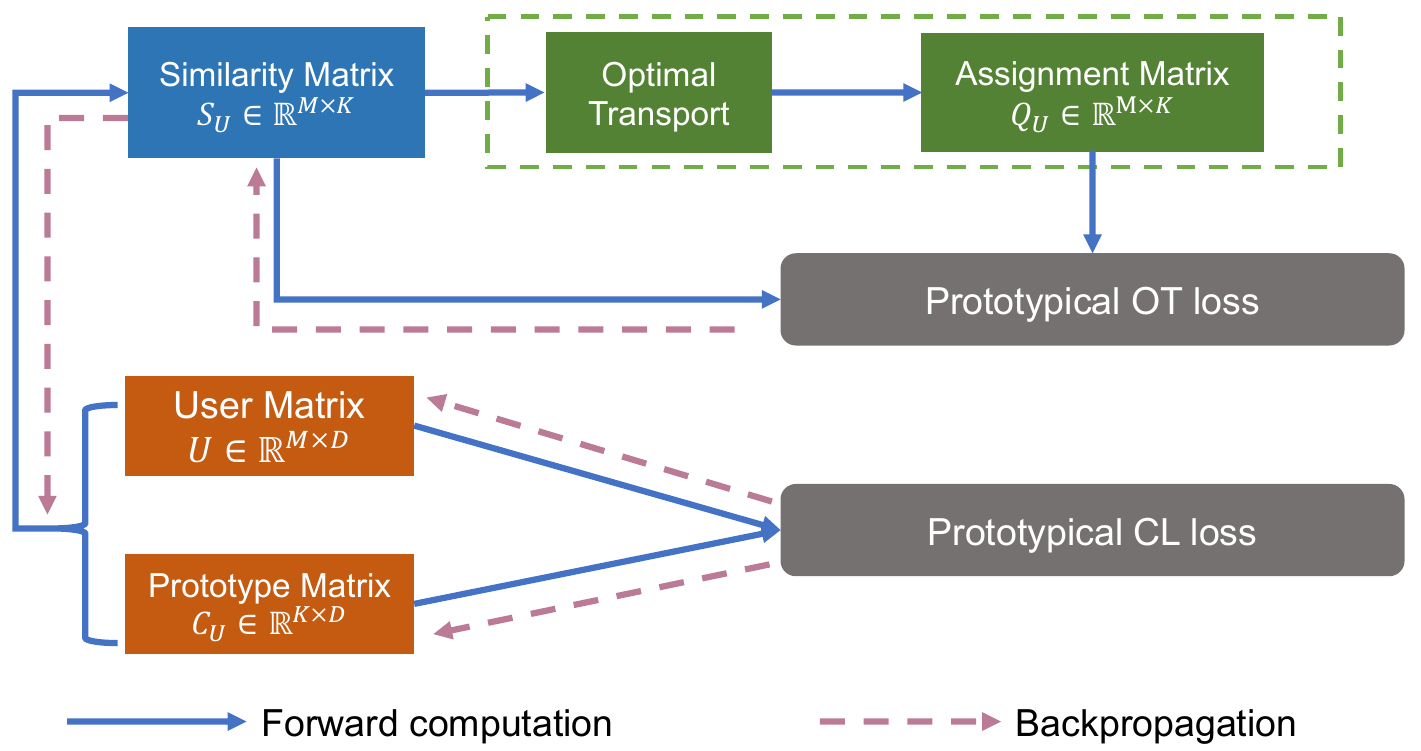}
  \caption{The training process of prototypical contrastive learning.}
  \label{pcl}
\end{figure}

The prototypical contrastive learning takes is built on the whole original node space (including both users and bundles) rather than in-batch data space. As shown in Figure \ref{pcl}, taking users as the example, let $U\in \mathbb{R}^{M\times D}$ be the user embedding matrix and we set a corresponding prototype matrix $C^U\in \mathbb{R}^{K\times D}$. If user representation $e_u$ is clustered to prototype $c_i^u$, the InfoNCE loss for users is:
\begin{equation}
    L_U^{proto}=\sum_{u\in U}-\text{log}\frac{\text{exp}(e_u\cdot c^u_i)/\tau}{\sum_{c^u_j\in C^U}\text{exp}(e_u\cdot c^u_j)}
\end{equation}

Analogously, the InfoNCE loss for bundles is:
\begin{equation}
    L_B^{proto}=\sum_{b\in B}-\text{log}\frac{\text{exp}(e_b\cdot c^b_i)/\tau}{\sum_{c^b_j\in C^B}\text{exp}(e_u\cdot c^b_j)}
\end{equation}

By summing the two contrastive losses, we obtain the prototypical contrastive loss:
\begin{equation}\label{eq:cl}
    L_{CL}^{proto}=L_U^{proto}+L_B^{proto}
\end{equation}

Now we present how to implement the prototype generation and learning. The basic criterion is to assign each user to the nearest prototype where the similarity is measured by dot product of their vectors. In practice, simply minimizing the InfoNCE loss under this naïve criterion suffers from the degeneracy problem since all data samples will be assigned to only a few prototypes. Hence we need to constrain that prototype assignment is equally partitioned \cite{nips20}. Let $S_U=UC_U^\top \in \mathbb{R}^{M\times K}$ be the similarity score matrix and $Q_U\in \mathbb{R}^{M\times K}$ denote the assignment matrix where each element $q_{up}$ is the probability that user $u$ is assigned to prototype $p$. The assignment of prototypes can be formulated as a relaxed Optimal Transport (OT) problem:
\begin{equation}\label{sk}
\begin{aligned}
    \text{max}_{Q_U\in \mathcal{Q_U}} \quad & \text{Tr}(S_UQ_U^\top)+\lambda H(Q_U) \\
    s.t. \quad & Q_U^\top\mathbf{1}_M=\frac{1}{K}\mathbf{1}_K \\
    & Q_U\mathbf{1}_K=\frac{1}{M}\mathbf{1}_M
\end{aligned}
\end{equation}
where $\mathbf{1}_M$ and $\mathbf{1}_K$ denote the vector of ones in dimension $M$ and $K$. The objective function is to maximize the total similarity between the vectors and the prototypes with an entropy regularization function $H(Q_U)=-\sum_{ij}Q_{U_{ij}}\text{log}Q_{U_{ij}}$ to control the smoothness of the assignment. The constraints guarantee that each prototype is associated with $\frac{M}{K}$ users.

According to Sinkhorn algorithm \cite{sink13,nips20}, the optimal solution $Q_U^*$ takes the form of a normalized exponential matrix:
\begin{equation}
Q_U^*=\text{Diag}(\boldsymbol{\alpha})\text{exp}\Big(\frac{S_U}{\lambda}\Big)\text{Diag}(\boldsymbol{\beta})
\end{equation}
where $\boldsymbol{\alpha} \in \mathbb{R}^K$ and $\boldsymbol{\beta} \in \mathbb{R}^B$ are renormalization vectors. We can solve it fast by means of fixed-point iteration. After getting the constrained prototype assignment $Q_U^*$, the loss function (\ref{eq:cl}) could be computed as:
\begin{equation}\label{loss:pr}
    L_{OT\_U}^{proto} = 	\left \langle Q_U, -\text{log} S_U	\right \rangle
\end{equation}
where $\left\langle \cdot \right\rangle$ is the Frobenius dot-product between two matrices and $\text{log}$ is applied element-wise. The loss (\ref{eq:cl}) encourages the assignment to be as uniform as possible. Similarly, we can get the loss $L_{OT\_B}^{proto}$ for bundles. Finally the loss of prototypical contrastive learning would be 
\begin{equation}
    L_{OT}^{proto} = L_{OT\_U}^{proto} + L_{OT\_B}^{proto}
\end{equation}

Note that computing $Q_U^*$ should not involve backpropagation (BP), while only the gradients of $U$ and $C$ are needed as illustrated in Figure \ref{pcl}. 

\subsection{Model Learning and Discussion}
\subsubsection{Model Learning}
The overall framework of GPCL is depicted in Figure \ref{over}. By combining all the aforementioned losses, for the $t_{th}$ sample from Gaussian embeddings, the loss would be:
\begin{equation}
    L_t=L_{BPR}+\gamma_{CL}* L_{CL}+\gamma_{PCL}*L_{CL}^{proto}+\gamma_{OT}*L_{OT}^{proto}
\end{equation}
where $\gamma_{CL}$, $\gamma_{PCL}$ and $\gamma_{R}$ are hyper-parameters to balance the four terms, and $L_{BPR}$ is the main loss. If we sample $T$ times, the total loss is 
\begin{equation}
    L=\sum_{t=1,2,...,T} \frac{1}{T}L_t 
\end{equation} 
\subsubsection{Uncertainty of Prediction}
Using Gaussian embeddings for nodes not only makes powerful representations, but also provides predictive uncertainty estimations \cite{sigir17}. Specifically, the final prediction score $y_{u,b}$ obeys a Gaussian distribution as well. Its mean and variance can be obtained by aggregating $T$ different prediction results. The value of the variance denotes the confidence of the model prediction. Exploration and exploitation strategy can be further adopted in real-world scenarios by considering the confidence. We can satisfy user's preference and expand the breadth of user's interest simultaneously, and boost up long-term utilities such as the social welfare.

\subsubsection{Original-Data-Level Cluster versus In-Batch-Level Cluster}
Usually the the cluster-based contrastive learning approach is implemented at in-batch level (after encoding), while we perform at original-data level (before encoding). We claim that in real-world recommendation scenarios, the number of underlying prototypes can be very large but the diversity of in-batch samples may be insufficient, thus enforcing the uniform assignment may lead to performance drop.




\section{EXPERIMENT}
In this section, we present the details of the experimental setups and the corresponding results, as well as an online A/B test, to answer the following questions:
\begin{itemize}
    \item \textbf{RQ1}: Does our proposed model GPCL achieve the best performance compared to other baselines?
    \item \textbf{RQ2}: What is the contribution of various components in our framework?
    \item \textbf{RQ3}: What is the effect of parameters change on the experiment? 
    \item \textbf{RQ4}: Can GPCL improve the performance of an existing model in a live recommender system?
\end{itemize}
\begin{table}[htbp]
  \caption{Statistics of three datasets.}
  \label{data}
  \begin{tabular}{cccc}
    \toprule
    DataSet&Youshu&NetEase&iFashion\\
    \midrule
    |User| & 8,039 & 18,528 & 53,897\\
    |Bundle| & 4,771 & 22,864 & 42,563\\
    |Item| & 32,770 & 12,3628 & 27,694\\
    |User-Item| & 138,515 & 1,128,065 & 2,290,645\\
    |User-Bundle| & 51,377 & 303,303 & 1,679,708\\
    |Bundle-Item| & 176,667 & 1,778,838 & 164,293\\
    Avg item interactions & 17.23 & 60.88 & 42.50\\
    Avg bundle interactions & 6.39 & 16.32 & 31.17\\
    Avg bundle size & 37.03 & 77.80 & 3.86\\
  \bottomrule
\end{tabular}
\end{table}
\begin{table*}[htbp]
        \caption{Results of both GPCL and baselines: the improvement is significant (p-value $\ll$ 0.01).} 
        \label{label:over_res}
        \resizebox{\textwidth}{!}{%
            \begin{tabular}{ccccccccccccc}
            \toprule
            \multirow{2}{*}{Method} & \multicolumn{4}{c}{Youshu}& \multicolumn{4}{c}{NetEase}& \multicolumn{4}{c}{iFashion}\\
            \cmidrule(r){2-5} \cmidrule(r){6-9} \cmidrule(r){10-13}  
            & rec@20 & ndcg@20 & rec@40 & ndcg@40 & rec@20 & ndcg@20 & rec@40 & ndcg@40 & rec@20 & ndcg@20 & rec@40 & ndcg@40 \\
            \hline
            \bfseries DAM       &0.2082 & 0.1198 & 0.2890 & 0.1418 & 0.0411 & 0.0210 & 0.0690 & 0.0281 & 0.0629 & 0.0450 & 0.0995 & 0.0579 \\
            \bfseries BundleNet &0.1895 & 0.1125 & 0.2675 & 0.1335 & 0.0391 & 0.0201 & 0.0661 & 0.0271 & 0.0626 & 0.0447 & 0.0986 & 0.0574\\
            \bfseries BGCN &0.2347 & 0.1345 & 0.3248 & 0.1593 & 0.0491 & 0.0258 & 0.0829 & 0.0346 & 0.0733 & 0.0531 & 0.1128 & 0.0671\\
            \bfseries CrossCBR &\underline{0.2801} & \underline{0.1669} & \underline{0.3781} & \underline{0.1941} & \underline{0.0794} & \underline{0.0431} & \underline{0.1206} & \underline{0.0539} & \underline{0.1149} & \underline{0.0885} & \underline{0.1653} & \underline{0.1062}\\
            \hline 
            \bfseries GPCL &{\bfseries 0.2882} & {\bfseries 0.1713} & {\bfseries 0.3963} & {\bfseries 0.2007} & {\bfseries 0.0833} & {\bfseries 0.0441} & {\bfseries 0.1270} & {\bfseries 0.0557} & {\bfseries 0.1216} & {\bfseries 0.0928} & {\bfseries 0.1756} & {\bfseries 0.1118}\\
            \bfseries \%Improv. &2.88\% & 2.62\% & 4.82\% & 3.38\% & 4.86\% & 2.42\% & 5.30\% & 3.33\% & 5.80\% & 4.85\% & 6.23\% & 5.27\%\\
            
            \bottomrule
            \end{tabular}%
        }
\end{table*}
\subsection{Experiment Setup}
\subsubsection{Datasets and Metrics}
To validate the effectiveness and robustness of our methods, we evaluate on three public datasets which are widely used for bundle recommendation:
\begin{itemize}
    \item \textbf{Youshu} \footnote{https://www.youshu.com/} is a book-review dataset \cite{dam}. The bundle is a book list that users may prefer.
    \item \textbf{NetEase} \footnote{https://music.163.com/} 
    is a music dataset \cite{netease}. The playlist composed of a set of songs acts as the bundle and can be favored/collected by users.
    \item \textbf{iFashion} is an outfit dataset where the outfit is made up of individual fashion items and treated as the bundle \cite{fashion}.
\end{itemize}

For a fair comparison, we adopt the same data preprocessing as existing research \cite{cbr}. The statistics of the datasets are shown in Table \ref{data}, where avg item interactions, avg bundle interactions, avg bundle size stand for the average number of interacted items per user, the average number of interacted bundles per user and the average number of affiliated items per bundle. To measure the validity of predicted preferences, we use Recall@n and NDCG@n as the metrics where $n\in \{20, 40\}$. 

\subsubsection{Competitors}
For a comprehensive evaluation, we compared our method with previous tailor-designed models for the bundle recommendation task:

\begin{itemize}
\item \textbf{DAM}. DAM \cite{dam} designs the attention mechanism to aggregate the item information of each bundle, and then jointly optimizes the user-bundle and user-item preference in a multi-task manner.
\item \textbf{BundleNet}. In BundleNet \cite{bnet}, the problem is formalized as a link prediction problem on a user-item-bundle tripartite graph, and is solved by a Relational Graph Convolutional Network model.
\item \textbf{BGCN}. BGCN \cite{bgcn} constructs two separate graphs, i.e., bundle-view graph and item-view graph. Then the GCN is used to learn representations, and they make predictions by fusing the representations from two views.
\item \textbf{CrossCBR}. Based on BGCN, CrossCBR \cite{cbr} uses LightGCN \cite{lgcn} to learn the representations from bundle-view graph and item-view graph separately, and then employs contrastive learning to model the cooperative association between the two views to improve performance. 
\end{itemize}
\subsubsection{Hyper-parameter and Training Settings}
Dimension of embedding vectors is 64. We optimize all models with Adam optimizer, where the learning rate is fixed at $10^{-4}$. The batch size is set to 2048. Sampling times $T$, $\gamma_{PCL}$ and $\gamma_R$ are additionally imported into our method, and they are explored by the grid search strategy with the range of \{1,2,4,6,8,10\}, \{0.01, 0.05, 0.1,0.15,0.2,0.25\} and \{0.01, 0.05, 0.1,0.15,0.2,0.25\}. For fairness, other hyper-parameters are consistent with the CrossCBR \cite{cbr} such as the temperature $\tau$ and data augmentation parameters. Each model is trained for 10 times and we record the model's average performance results. All the models are trained using Pytorch 1.7, NVIDIA V100 GPUs.

\subsection{Overall Performance (RQ1)}
Table \ref{label:over_res} presents the performance that each approach obtains. Boldface denotes the best performance
and underline indicates the strongest result of the baselines. 
We compare GPCL with the baselines. \%Improv. measures the relative improvements of GPCL over the strongest result of the baselines. We have the following observations:

Graph-based models generally exhibit better results, proving the expressive power of graph learning for the bundle recommendation task. However, among the graph-based models, BundleNet performs poorly, and we ascribe this to the failure of differentiating users' behavioral similarity and bundles' content relatedness. Through modeling user preference in two separate views, BGCN and CrossCBR achieve significant gains. Aided by contrastive learning, CrossCBR reaches SOTA performance. Our GPCL can beat the SOTA, i.e., CrossCBR, on all the three datasets. This demonstrates the effectiveness and robustness of our method as well as the key innovative components we propose. In the following sections, we will present more detailed analysis with regard to the proposed components.

\subsection{Ablation study (RQ2)}
To evaluate the proposed components of GPCL, we conduct a list of ablation studies as reported in Table \ref{table:aba-res} where \%Improv. measures the relative improvements over CrossCBR. Due to the space limitation, we only present rec@20 and ndcg@20 on Youshu and NetEase datasets. 
\subsubsection{Effectiveness of Gaussian Embedding Module}
In order to verify whether the Gaussian Embedding Module contributes to the performance, we first remove the prototypical contrastive learning module from the model, named Gauss-Emb. 
\begin{table}[htbp]
        \caption{Results of ablation methods.} 
        \label{table:aba-res}
            \begin{tabular}{ccccc}
            \toprule
            \multirow{2}{*}{Method} & \multicolumn{2}{c}{Youshu}& \multicolumn{2}{c}{NetEase}\\
            \cmidrule(r){2-3} \cmidrule(r){4-5}
            & rec@20 & ndcg@20 & rec@20 & ndcg@20 \\
            \hline
            \bfseries CrossCBR &0.2801 & 0.1669 & 0.0794 & 0.0431 \\
            \hline 
            \bfseries CrossCBR-2D &0.2810 & 0.1674 & 0.0792 & 0.0430 \\
            \hline 
            \bfseries Proto-Batch & 0.2785 & 0.1658 & 0.0791 & 0.0430\\
            \hline 
            \bfseries Gauss-Emb &0.2828 & 0.1680 & 0.0823 & 0.0438 \\
            \bfseries \%Improv. &0.98\% & 0.69\% & 3.60\% & 1.62\%\\
            \hline 
            \bfseries Proto-CL &0.2870 & 0.1702 & 0.0809 & 0.0438\\
            \bfseries \%Improv. &2.45\% & 1.95\% & 1.83\% & 1.62\% \\
            \hline 
            \bfseries GPCL &{\bfseries 0.2882} & {\bfseries 0.1713} & {\bfseries 0.0833} & {\bfseries 0.0441}\\
            \bfseries \%Improv. &2.88\% & 2.62\% & 4.82\% & 2.42\% \\

            \bottomrule
            \end{tabular}%
\end{table}
Gauss-Emb shows a performance decline compared with GPCL but still outperforms the baseline CrossCBR, demonstrating the effectiveness of modeling uncertainty.

When Gaussian embedding is introduced, each node is represented by two $D$-dimensional vectors (i.e., the mean and variance vectors), making the the number of representation parameters doubled. For a fair comparison, we design an ablation study to explore the influence of the parameter size. We set the embedding size of CrossCBR to $2D=128$ during training and obtain CrossCBR-2D. From Table \ref{table:aba-res} we observe that on Youshu dataset, the performance of CrossCBR-2D is just slightly better than CrossCBR while even underperforms CrossCBR-2D on NetEase dataset, indicating that simply increasing the parameter size has little effect with respect to the performance.


To demystify the working mechanism behind Gaussian embedding, we further evaluate our framework particularly against the nodes (users and bundles) with uncertainties. One of main sources of uncertainties is the marginal (nodes with few past engagements) nodes, because very little information about their information can be obtained from historical actions. 
Therefore we divide the nodes into four groups by their frequencies and exam the uncertainty captured by our method. The uncertainty is measured by the learned mean and variance vector and calculated as follows:
$$uncertainty=\sum_{i=1,2,...,D}\frac{\sqrt{\sigma'_i}}{|\mu_i|}$$
where $|\cdot|$ represents the absolute value; $\sigma'_i$
denotes the $i^{th}$ value of the transformed variance vector and $|\mu_i|$ denotes the the $i^{th}$ value of the mean vector. 
\begin{table}[htbp]
  \caption{Uncertainty of nodes with different frequency.}
  \label{tabel:freq}
    \begin{tabular}{ccccc}
    \toprule
        Dataset& User-freq & Uncertainty & Tag-freq & Uncertainty \\
        \midrule
        \multirow{4}{*}{Youshu} & 1-10& 0.4805& 1-10& 27.11\\
          & 11-30& 0.2413& 11-50& 6.85\\
         & 31-50& 0.1642& 51-200& 4.06\\
         & 51-& 0.0787& 201-&1.26\\
         \hline
         \multirow{4}{*}{NetEase}& 1-10& 0.6585& 1-10& 128.0\\
         & 11-100& 0.4245& 11-100& 28.57\\
         & 101-200 & 0.1353 & 101-200  & 6.79\\
         & 201-  & 0.0796      & 201-     & 6.24 \\
        \bottomrule
    \end{tabular}
\end{table}
As shown in Table \ref{tabel:freq}, the model assign larger uncertainties to the low-frequency nodes, which suggests that Gaussian embeddings capture the confidence of the nodes.

\subsubsection{Effectiveness of Prototypical Contrastive Learning Module (RQ3)}
To evaluate the effectiveness of the prototypical contrastive learning module, we design Proto-CL by removing Gaussian embedding from GPCL. As expected, Proto-CL is still better than CrossCBR by a large margin. In conclusion, both modules, i.e., Gaussian embedding and prototypical contrastive learning, collaboratively contribute to the performance.

Recall that our prototypical contrastive learning is trained on whole original nodes. In order to justify the effectiveness of this learning paradigm, we make a comparison with Proto-Batch where the prototypical contrastive learning is deployed at in-batch level. As reported in Table \ref{table:aba-res}, Proto-Batch underperforms baseline CrossCBR on both datasets, which indicates that the prototypical contrastive learning at whole original node space is necessary to capture the complete contextual information.
\subsection{Hyper-parameter study}
Our proposed GPCL framework has introduced several vital hyper-parameters, which may affect performances during training. In this section, we investigate the importance and sensitivity of these hyper-parameters while fixing the others.

\subsubsection{Influence of the number of prototypes}
\begin{figure}[htbp]
	\begin{center}
		\begin{minipage}{0.45\linewidth}
			\includegraphics[width=\textwidth]{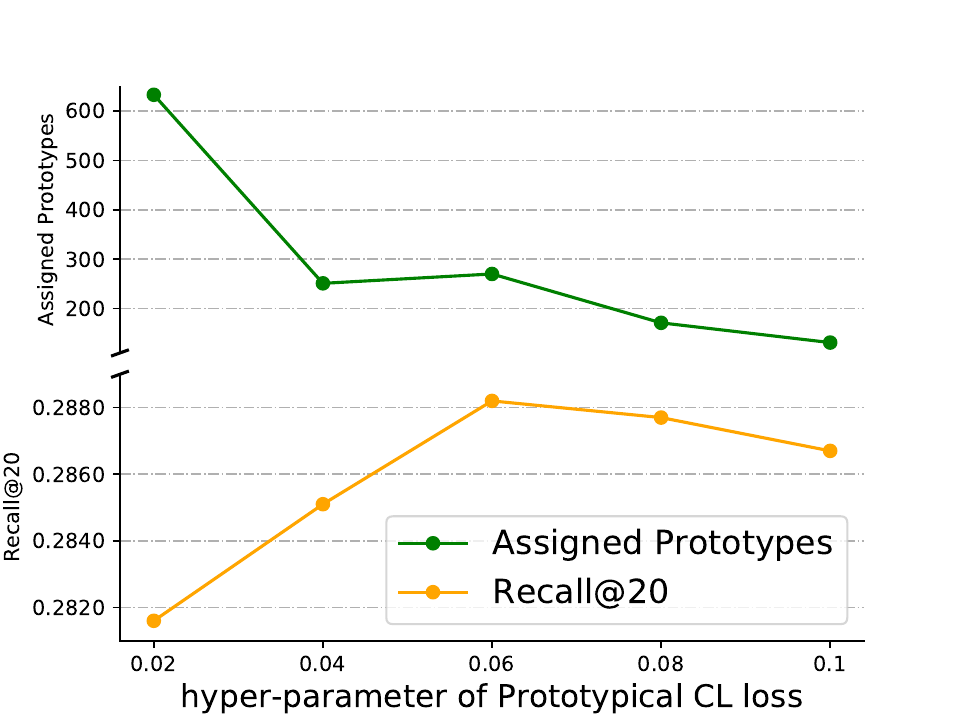}\\
			\centering{(a) Youshu}
		\end{minipage}
		\begin{minipage}{0.45\linewidth}
			\includegraphics[width=\textwidth]{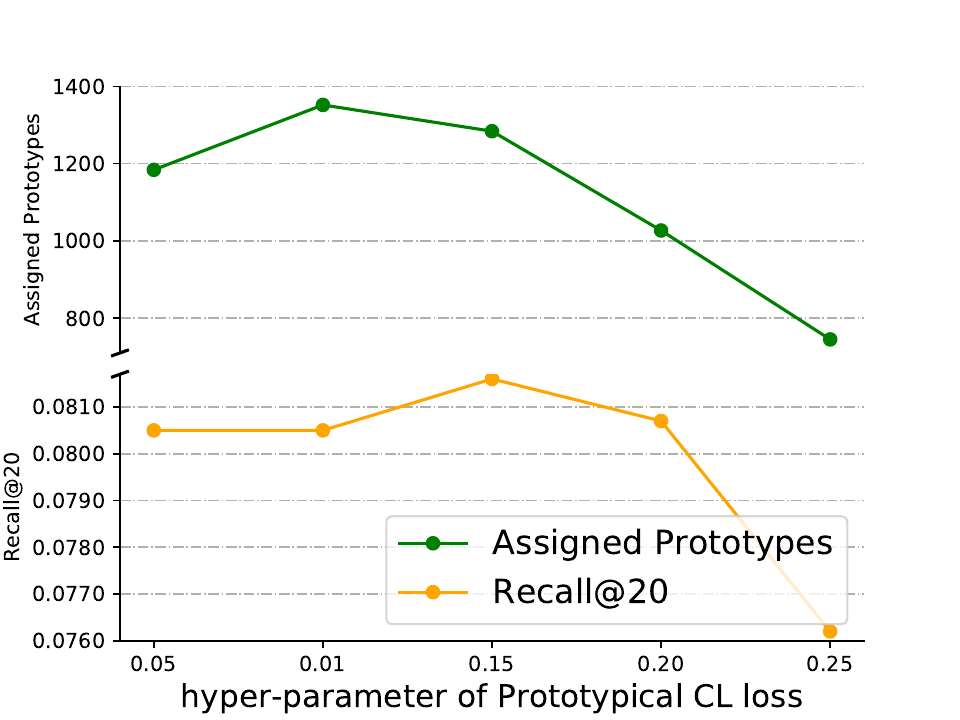}\\
			\centering{(b) NetEase}
		\end{minipage}      
	  
		\caption{Sensitivity analysis for the number of prototypes $T$.}    
            \label{fig:proto-pc}
	\end{center}
\end{figure}

The prototypical contrastive learning module involves two loss $L_{CL}^{proto}$ and $L_{OT}^{proto}$, where the former is the prototypical InfoNCE loss and the latter encourage the uniform assignment on the prototypes. The corresponding two weights $\gamma_{PCL}$ and $\gamma_{OT}$ determine the number of actually assigned prototypes. We evaluate the impact of prototype number by fixing $\gamma_{OT}$ and varying $\gamma_{CL}$. Figure \ref{fig:proto-pc} shows the number of assigned prototypes and the performance (Recall@20 due to space limitations) with the value of $\gamma_{PCL}$. It can be observed that either too small or too large $\gamma_{PCL}$ leads to performance drop and large $\gamma_{PCL}$ could reduce utilization of prototypes. We conjecture that the model degenerates to the case without prototypes (i.e., each node acts as a prototype itself) when $\gamma_{PCL}$ is too small, and too large $\gamma_{PCL}$ brings about assignment collapse (i.e. all nodes are assigned to few prototypes). 

\vspace{-0.2cm}

\subsubsection{Influence of Sampling Size}
\begin{figure*}[htbp]
	\begin{center}
		\begin{minipage}{0.282\linewidth}
			\includegraphics[width=\textwidth]{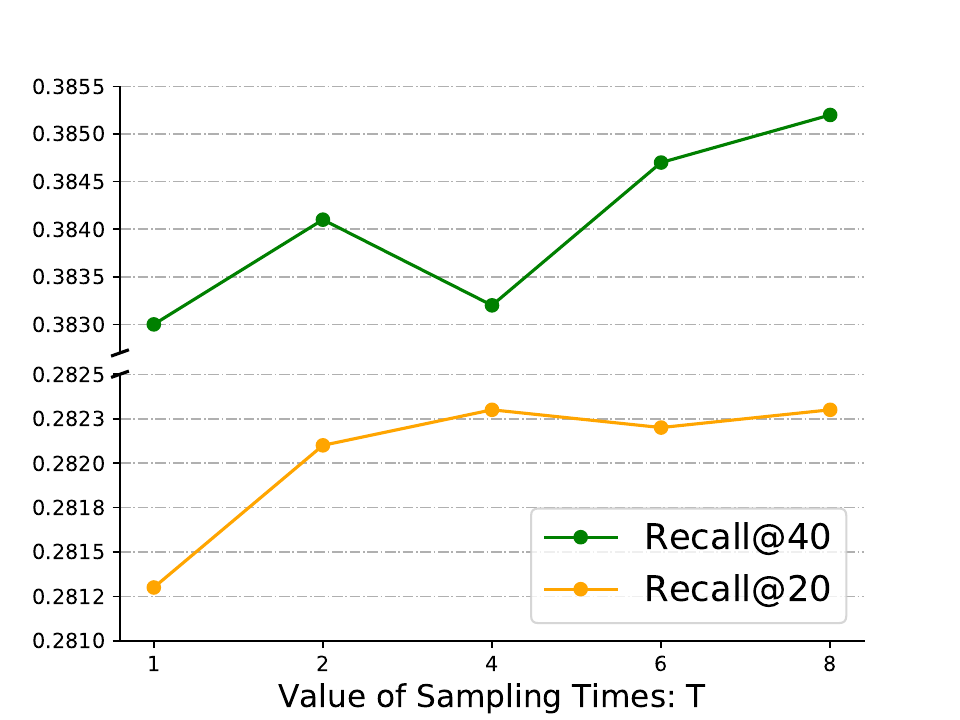}\\
			\centering{(a) Youshu\_Recall}
		\end{minipage}
		\begin{minipage}{0.282\linewidth}
			\includegraphics[width=\textwidth]{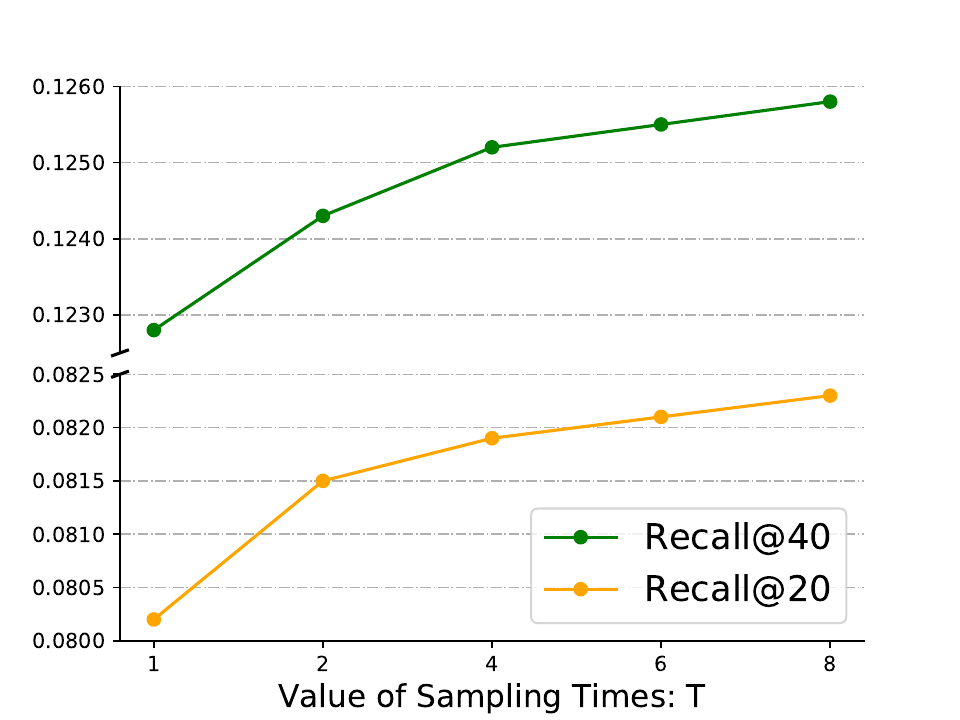}\\
			\centering{(b) NetEase\_Recall}
		\end{minipage}      
		\begin{minipage}{0.282\linewidth}
			\includegraphics[width=\textwidth]{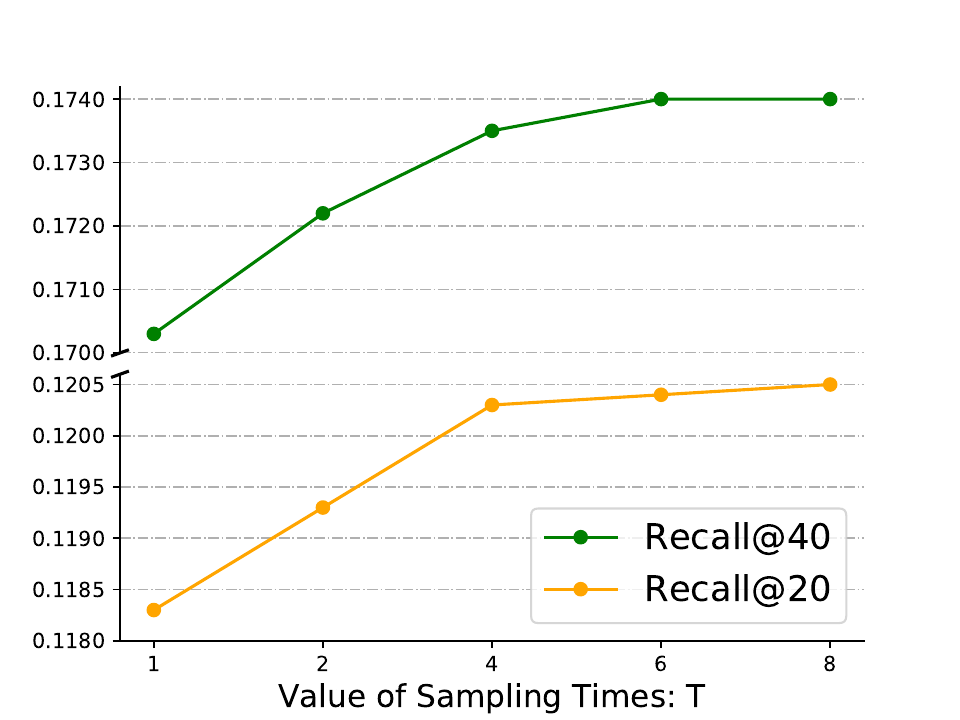}\\
			\centering{(c) iFashion\_Recall}
		\end{minipage}
		
		\begin{minipage}{0.282\linewidth}
			\includegraphics[width=\textwidth]{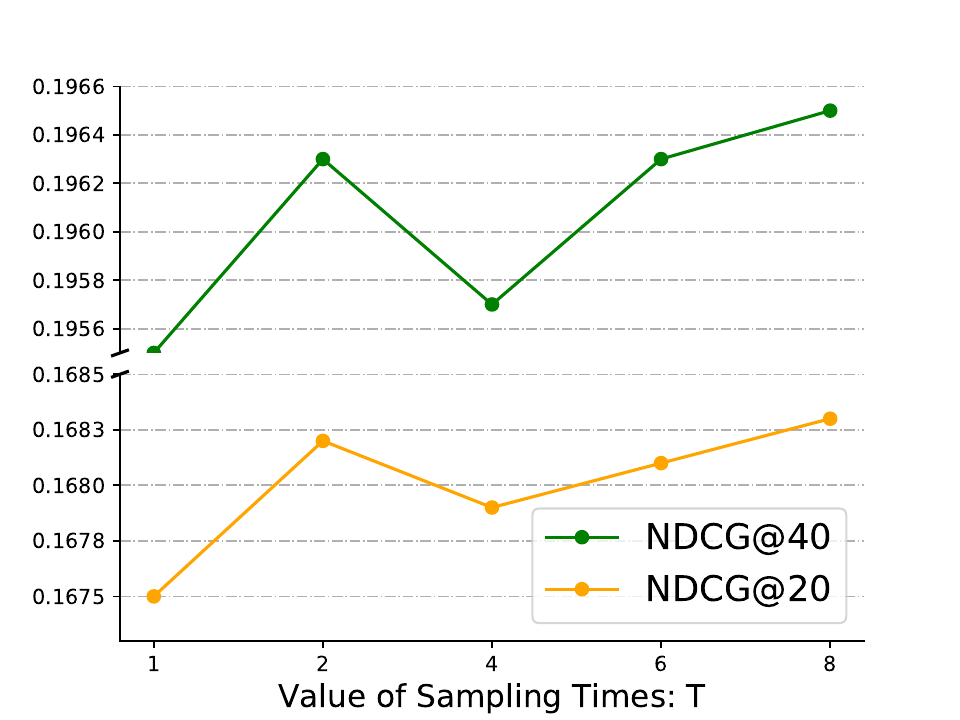}\\
			\centering{(d) Youshu\_NDCG}
		\end{minipage}
		\begin{minipage}{0.282\linewidth}
			\includegraphics[width=\textwidth]{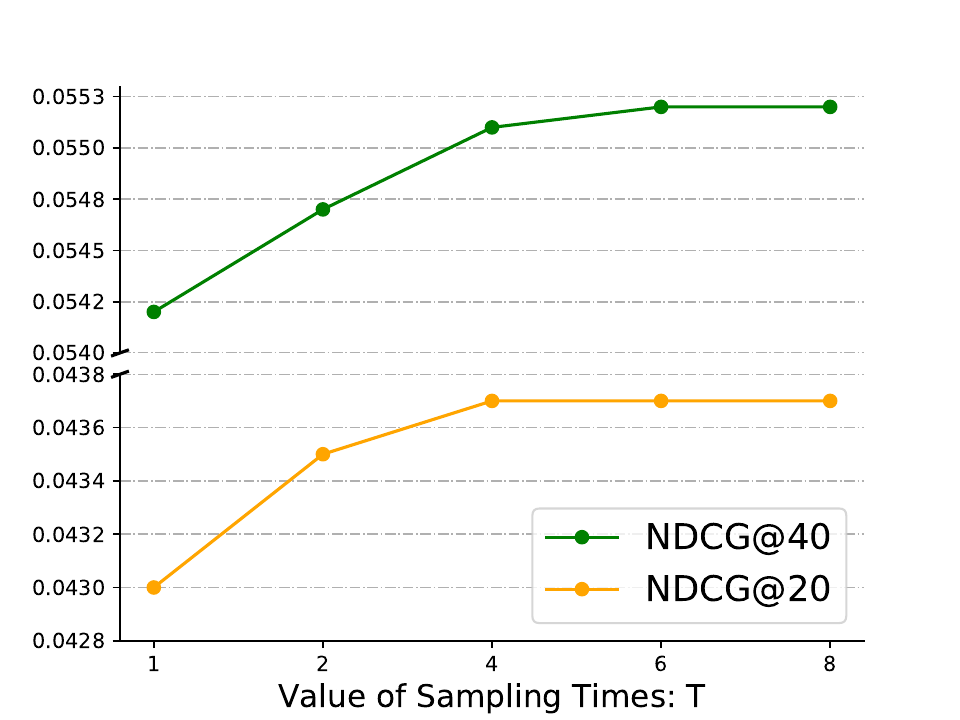}\\
			\centering{(e) NetEase\_NDCG}
		\end{minipage}      
		\begin{minipage}{0.282\linewidth}
			\includegraphics[width=\textwidth]{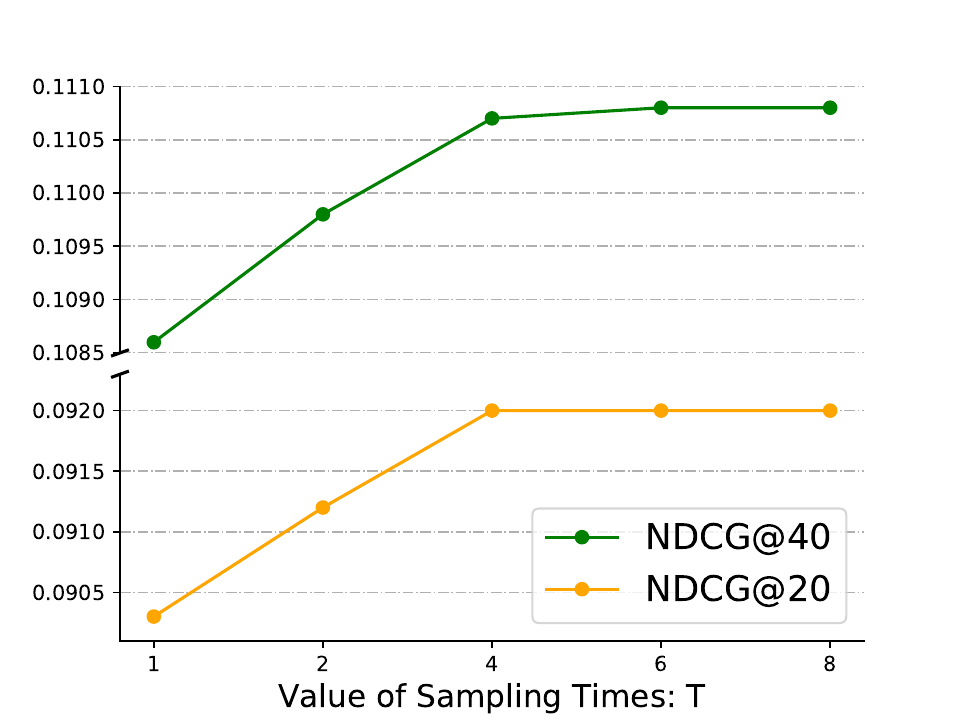}\\
			\centering{(f) iFashion\_NDCG}
		\end{minipage}
		\caption{Sensitivity analysis for sample size $T$.}   
            \label{fig:gauss_k}
	\end{center}
\end{figure*}

Our proposed Gaussian embedding module has involved a hyper-parameter: the sampling times $T$. To investigate its importance and sensitivity, we set $T$ from 1 to 8 and the results are shown in Figure \ref{fig:gauss_k}. From the line chart, we can observe that as the sampling size grows, the performance improves before the model reaches a state of complete convergence and then remains stable when $T$ is large.

\subsection{Online A/B test (RQ4)}
To verify the actual benefits of our model, we conduct online A/B test in the real-world production environment. Our experiments are based on the log data from a certain industry of the Alibaba's e-commerce platform. The data covers 2,238,947 users, 2,682,634 items and 7302 bundles where the average number of affiliated items per bundle is 182. The task is to select proper bundles for the users to satisfy their preferences. Our goal is to maximize the total number of paying customers, and we use conversion rate (CVR) and click-through\&conversion rate (CTCVR) as the metrics to evaluate the performance of the online A/B test. 

Due to the limited resources, we only compare our best model GPCL with the current solution working online. For the control group, we adopt the current online solution, a simple yet effective strategy where the user-bundle preference score is calculated by some tailor-designed rules. It has been deployed for a long time due to its high efficiency and stable performance. For the experimental group, the preference score is produced by GPCL. The numbers of users in the control and experimental groups are approximately identical, each accounting for 50\%.

\begin{table}[htbp]
\caption{Online A/B Test Results.}
\label{table:online}
    \begin{tabular}{ccc}
        \toprule
        Metric & CVR     & CTCVR   \\
        \hline
         Lift rate      & 2.03\%      & 2.06\%  \\
        \bottomrule
    \end{tabular}%
\end{table}

Table \ref{table:online} illustrates the lift rate of GPCL over the baseline with regard to the two metrics. We could see that GPCL achieves a significant improvement. The results demonstrate the effectiveness of our proposed model.

\section{RELATED WORK}
\textbf{Bundle recommendation}. DAM \cite{dam} first design an attention network to aggregate the item embeddings in a bundle in a multi-task manner to jointly optimize the user-bundle and user-item preference, but fail to explore the affiliation between items and bundles. Recent evidence \cite{cbr,bnet,bgcn} suggests that graph representation learning is an effective technique to capture the complicated topology and higher-order connectivity of user-bundle-item. BundleNet \cite{bnet} builds a unified tripartite graph and use GNNs \cite{rgcn} to perform representation learning which can implicitly incorporate the intermediate role of items between users and bundles. However, the cooperation of bundle view and item view are not well differentiated. Recently CrossCBR \cite{cbr} further involve contrastive learning based on BGCN and achieve SOTA performance.

\textbf{Gaussian embedding of graphs}. KL-divergence-based methods \cite{deepGE, dmkd} to measure the similarity of graph nodes are widely used. DeepGE \cite{deepGE} propose an unsupervised approach where the mean and variance vectors of each node are obtained by a deep encoder, and they utilize the node's neighbor information to minimize the KL-divergence distance. DMKD \cite{dmkd} employ a given node similarity metric to measure the global structural information, then generate structural context for nodes and finally learns node representations via Gaussian embedding. GE \cite{GE} propose an end-to-end framework for the large-scale graph where they learn from both node attributes and graph structural information. Another line of research directly applys Gaussian embedding on nodes without considering graph structure \cite{sigir17,ijcai19}. GECF \cite{sigir17} represent each node by a normal-gamma distribution where the parameters are learned through the rank loss. GeRec \cite{ijcai19} assume that the representation of each node follows a Gaussian distribution and then sample several times as the input tensors of the subsequent CNN encoder.

\textbf{Clustering-based contrastive learning}. Due to the sampling bias issue of instance-wise contrastive learning, cluster-based contrastive learning is proposed in recent studies \cite{ncl,nips20,proto-1,cluster-1} where they contrast between instances and cluster centroids to capture underlying semantic information. In their work \cite{ncl, pcl}, cluster centroids are obtained by k-means, and optimized by an Expectation-Maximization framework. While some other methods parameterize the cluster prototypes directly and update them with model in each step. For the computer vision task, SwAV \cite{nips20} introduce the concept of prototypes which can be seen as the context (i.e., a group of semantic neighbors) of each representation. They use constrained optimization algorithm and cross-entropy loss to guide the learning of prototype vectors and the assignment of instances. Similarly, PGCL \cite{pgcl} adopt the prototype-based method in the graph classification task where they design a reweighted contrastive loss where negatives having moderate prototype distance enjoy relatively large weights to mitigate sampling bias issue.

\section{CONCLUSION AND FUTURE WORK}
In this paper, we present a Gaussian Graph with Prototypical Contrastive Learning framework for the bundle recommendation task. First, we propose a Gaussian embedding module to model the uncertainty. To capture the contextual information and obtain more refined representations, we further develop the prototypical contrastive learning module. Extensive experiments demonstrate that benefiting from the proposed components, we achieve new state-of-the-art performance compared to previous methods on three public datasets. The online A/B test exhibits the superiority of our method in the real-world production environment. As future work, we are very interested in the downstream exploration and exploitation strategy by considering the uncertainty, to improve users' shopping experience by enhancing novelty and serendipity at little cost.


\normalem
\bibliography{sample-base}


\end{document}